\documentclass[english,12pt,twoside]{article}
\usepackage{amssymb}
\usepackage{amsmath}
\usepackage[pdftex]{graphicx}
\newtheorem{df}{Definition}
\usepackage{pst-node}
\begin{document}

\date{}
\title{Decisions in elections --- transitive or intransitive quantum preferences}
\author{Marcin Makowski\footnote{makowski.m@gmail.com}, Edward W. Piotrowski\\[1ex]\small
Institute of Mathematics, University of Bia\l ystok,\\\small
Akademicka 2, PL-15424, Bia{\l}ystok,
Poland}
\maketitle\begin{abstract}
 Our preferences depend on the circumstances in which we reveal them. We will introduce a dependency which allows us to illustrate the relation between the possibility of winning of particular candidates in a quantum election and the type of preference.  It occurs that if voters start to clearly prefer one of the candidates, the significance of intransitive preferences in the quantum model decreases.  This dynamic change cannot be observed in the case of the classical model.
\end{abstract}
\section{Introduction}
Any relation $\succ$ existing between the elements of a certain set is called \emph{transitive} if $A\succ C$ results from the fact that $A\succ B$ and $B\succ C$ for any three elements  $A$, $B$, $C$. If this condition is not fulfilled then the relation will be called \emph{intransitive} (not transitive). \\
The axiom concerning the intransitiveness of preference relations is one of the basic assumptions of the choice theory. It is often identified with the rationality of the taken decisions.
 One of the main arguments put forward by many experts, proving the irrationality of preferences which violate transitivity, is the so-called ,,money pump'' (see \cite{r2}). \\
Despite the fact that intransitivity appears to be contrary to our intuition, life provides many examples of intransitive orders. Rivalry between species may be intransitive. For example, in the case of fungi, Phallus impudicus replaced Megacollybia platyphylla, M. platyphylla replaced Psathyrella hydrophilum, but P. hydrophilum replaced P. impudicus \cite{r15}. Similarly, the experiment can be explained on bees, which make intransitive choices between flowers \cite{r16}.
The best known and socially significant example of intransitivity is Condorcet's voting paradox. Consideration of this paradox led Arrow (in the 20th century) to prove the theorem that there is no procedure of successful choice that would meet a democratic assumption \cite{r24}. Interesting examples of intransitivity are provided by probability models (Efron's dice  \cite{r34}, Penney's game \cite{r35}).\\
Intransitive strategies are sometimes a consequence of negligent, rash decisions. They also appear when we are not able to conduct an accurate valuation of the assessed goods. If we lack information indispensable in a conscious decision--making process (we are indecisive). Often the criteria which decide about a choice refer to a sphere of feelings which are difficult to compare. 
Ulam makes a mention of it in his autobiography \cite{ru}. He tried to evaluate the taste of fruits and he claimed that this relation is intransitive (plums could be better than nuts, which were better than apples, but apples were better than plums). 
The division into transitive and intransitive strategies (preferences) does not necessarily have to be equal to the division into rationality and irrationality. It may, however, reflect the process of decision-making in conditions of uncertainty or the lack of decisiveness of the chooser. Intransitive strategies are still a mysterious feature of the human thinking process and the understanding of the reasons behind the presence of intransitivity may be of use so as to better understand thinking mechanisms and for the research concerning modelling of artificial intelligence. This concept is worth analysing also within the context of tools which are provided by the quantum game theory which has recently been developing intensively \cite{r4,r5, r6,  r7,r26, r8, r9, r10}, where the formalism of Hilbert's space is applied for constructing decision--making algorithms. This non--classical approach may lead to interesting and qualitatively new effects which cannot be achieved by means of traditional probabilistic models.
\vspace{0.3cm}\\
In this article we will once again take a look at a very simple game analysed in \cite{r1} and a different way of modelling this game known from the previous analyses  \cite{r2op, r2}. We are going to pay attention to an interesting relationship (occurring in the quantum model) between intransitive strategies and the increase of the player's certainty (he starts to be in favour of one of the available alternatives). We will show that if voters start to clearly prefer one of the candidates, the significance of intransitive preferences in the quantum model decreases.This seems to be a fairly natural and intuitive characteristic, especially if intransitive preferences are interpreted as an expression of uncertainty in the decision making process. Analysis of a well--known example (Condorcet paradox) will allow us to obtain such a result.\\
 Suppose we have three candidates, $A$, $B$, and $C$. Let's assume that each voter's preference is independently selected from these (transitive) preferences:  
\begin{enumerate}
	\item $A\succ B\succ C$\,
	\item $B\succ C\succ A$\,
	\item $C\succ A\succ B$\,,
\end{enumerate}
 with probability $w_1$, $w_2$, $w_3$, respectively. First we assume  that $w_1=w_2=w_3=1/3$.  Collective preference is intransitive: with probability $2/3$  $A\succ B$ (from 1. and 3.), with probability $2/3$ $B\succ C$ (from 1. and 2.) and with probability $2/3$ $C\succ A$ (from 2. and 3.). 
The significance of the collective intransitive preference will reduce along with reduction of probabilities $w_2$ and $w_3$ (reduction of probability that $C\succ A$ ) and increase of probability  $w_1$ (i.e., voters start to clearly prefer candidate $A$). In the Condorcet paradox, pure strategies, i.e. transitive preferences, are mixed. This model therefore favours transitive preferences, and thus cannot be treated as an effective tool in solving the conflict between transitive and intransitive preferences with all related implications.  This article refers to a model devoid of this shortcoming and takes both intransitive and transitive preferences into equal account.
Illustrating the problem we will relate to an election interpretation of the game proposed in  \cite{r2}. It will allow us to situate the problem in a real--life context. We will analyse the influence of the increase of support (understood as a probability of winning the election) for one of the candidates for the significance of intransitive preferences. 

Owing to the quantum approach, the article naturally singles out the class of relevant intransitive stratiegies which has not been subject to prior analysis.
\begin{df}
The intransitive strategy will be called the relevant strategy, if there is no transitive strategy of the same consequences with the same assumption.
\end{df}
Relevant intransitive strategies may occur only in the quantum game model and their significance decreases proportionately to the player's increase in determination to make a given choice. In everyday life, we often tend to be perfectly sure about our decisions. On the other hand, however, we are equally often insecure about them (we are not absolutely convinced that we have made a good choice).  It is extremely difficult to construct a mathematical model describing such relations, one which would fully render their complexity. Therefore, it must be emphasised that the conclusions of this dissertation relate only to a simple model of behaviour, and it is difficult to say whether they would be suitable in the case of more complex models. In the article we refer to the well--known game model. Simple modification of its parameters allows for an interesting characteristic to be observed concerning increase in the player's determination (which is defined as the probability of deciding on one of the possible options). Mathematical rules employed for the purposes of this article have been introduced (partially) in paper \cite{r2}. Significant terms are explained in the next paragraph in order to facilitate reading and make the article complete.  

\section{The model}
Let us assume that three candidates (no 0, no 1, no 2) take part in the election. 
We do not refer to any concrete election procedure here, however, we assume that the elections ensue in two stages. In the first stage there is an elimination of one of the candidates in order to make a final choice in the second stage (of one of the two remaining candidates).
We can observe such a construction of election in many countries (mainly the European ones) where two candidates with the highest social support enter the second stage (the so called second round of the elections). In this type of elections the candidate's chances depend also on the strength of the candidate with whom he will have to compete in the second round. It may occur that the candidate with the largest support in the first round will lose in the second round (the choice depends on the context; we often vote ''against'' and not ''for'').  \newline
Let us introduce the symbols.
Let $q_i$ denote the probability that the $i$ candidate will not enter the second round. 
We denote by $P(C_k|B_j)$ the probability of choice of candidate $k$ in the second round when the decision concerned a pair of candidates in which candidate $j$ is not present. The probability of the $k$ candidate's victory will be denote by $\omega_k$:
\begin{equation}\label{zwiazek}
\omega_k=\sum_{j\neq k}P(C_{k}|B_{j})\,q_j,\,\,\,\,\, j,k=0,1,2\,.
\end{equation}
Any six conditional probabilities ($P(C_1|B_0)$, $P(C_2|B_0)$, $P(C_0|B_1)$, $P(C_2|B_1)$, $P(C_0|B_2)$, $P(C_1|B_2)$) that for a fixed triples ($q_0,q_1,q_2$) and ($\omega_0,\omega_1,\omega_2$) fulfill (\ref{zwiazek}) will be called a \emph{voters optimal strategy} (``the collective voter'' --- the electorate). This strategy could come
into being as a superposition of the strategies of single voters interfering
with each other (it is pure strategy).\\
After elementary modification we introduce the following relation:
\begin{eqnarray}
\label{solution}
q_0=\frac{1}{d}\bigg(-P(C_2|B_1)\,\omega_0+P(C_2|B_1)P(C_0|B_2)+\nonumber\\ +(1-P(C_2|B_1)-P(C_0|B_2))\,\omega_2\negthinspace \bigg),\nonumber\\
q_1=\frac{1}{d}\bigg(-P(C_0|B_2)\,\omega_1+P(C_0|B_2)P(C_1|B_0)+\\ +(1-P(C_0|B_2)-P(C_1|B_0))\,\omega_0\negthinspace\bigg),\nonumber\\
q_2=\frac{1}{d}\bigg(-P(C_1|B_0)\,\omega_2+P(C_1|B_0)P(C_2|B_1)+\nonumber\\ +(1-P(C_1|B_0)-P(C_2|B_1))\,\omega_1\negthinspace\bigg).\nonumber\\\nonumber
\end{eqnarray}
The above relation defines a mapping $\mathcal{A}:D_3\rightarrow T_2$ of the three--dimensional cube ($D_3$) into a triangle ($T_2$) two--dimensional simplex 
 ($q_0+q_1+q_2=1$ and $q_i\geq 0$), where $d$ is the determinant of the matrix of
parameters $P(C_j|B_k)$ (see \cite{r1}). The barycentric coordinates of a point of
this triangle are interpreted as probabilities $q_0, q_1$ and $q_2$.  

For further deliberations (in quantum case) we will assume the construction of conditional probabilities $P(C_k|B_j)$  proposed in work \cite{r2op}. 
It is based on the replacement of a cube $D_3$  with a sphere $S_2$ with the use of a probabilistic interpretation of vector coordinates in a two--dimensional Hilbert space $H_2$ and the concept of the so called conjugated bases which have already played an essential role e.g. in quantum cryptography \cite{r27} and quantum market games \cite{r28}. The coordinates of the same strategy
$| z \rangle\negthinspace \in H_2$\, read (measured) in various bases define voter’s
preferences toward a candidate pair represented by the base vectors. Squares of their moduli, after normalization, measure the conditional
probability of voter’s making decision in choosing a particular candidate, when the choice is related to the suggested
candidate pair. Ultimately the probabilities $P(C_{k}|B_{j})$ which are of interest to us have the following form \cite{r2op}:
 \begin{eqnarray}
P(C_0|B_2)=\frac{1-x_3}{2}, \qquad P(C_1|B_2)=\frac{1+x_3}{2},\nonumber \\
P(C_0|B_1)=\frac{1+x_1}{2},\qquad P(C_2|B_1)=\frac{1-x_1}{2}, \\
P(C_1|B_0)=\frac{1+x_2}{2},\qquad P(C_2|B_0)=\frac{1-x_2}{2},\nonumber  
	\end{eqnarray}
where $(x_1,x_2,x_3)\in S_2$.\newline
Combination of the above projection with (\ref{solution}) results in the projection $\mathcal{A}_q:S_2\rightarrow T_2$, of two--dimensional sphere $S_2$ into a triangle $T_2$.
\newline
In case of random selections we may talk about order relation:
\begin{displaymath}
\rm{candidate\,\, no}\,\,0 < \rm{candidate\,\,no} \,\,1\,,
\end{displaymath}
when from pair $(0, 1)$ we are willing to choose the candidate no 1 ($P(C_{0}|B_{2})<P(C_{1}|B_{2})$).
We deal with an intransitive choice (strategy) if one of the following conditions is fulfilled:
 \begin{itemize}
  \item $P(C_2|B_1)<\frac{1}{2}$, 
  $P(C_1|B_0)<\frac{1}{2}$, 
  $P(C_0|B_2)<\frac{1}{2}$\,,
  \item $P(C_2|B_1)>\frac{1}{2}$, 
  $P(C_1|B_0)>\frac{1}{2}$, 
  $P(C_0|B_2)>\frac{1}{2}$\,.
 \end{itemize}

\section{Decrease of importance of intransitives strategies} 
\begin{figure}[htbp]
          \centering{
        \includegraphics[
           height=1.55in,
           width=1.65in]%
          {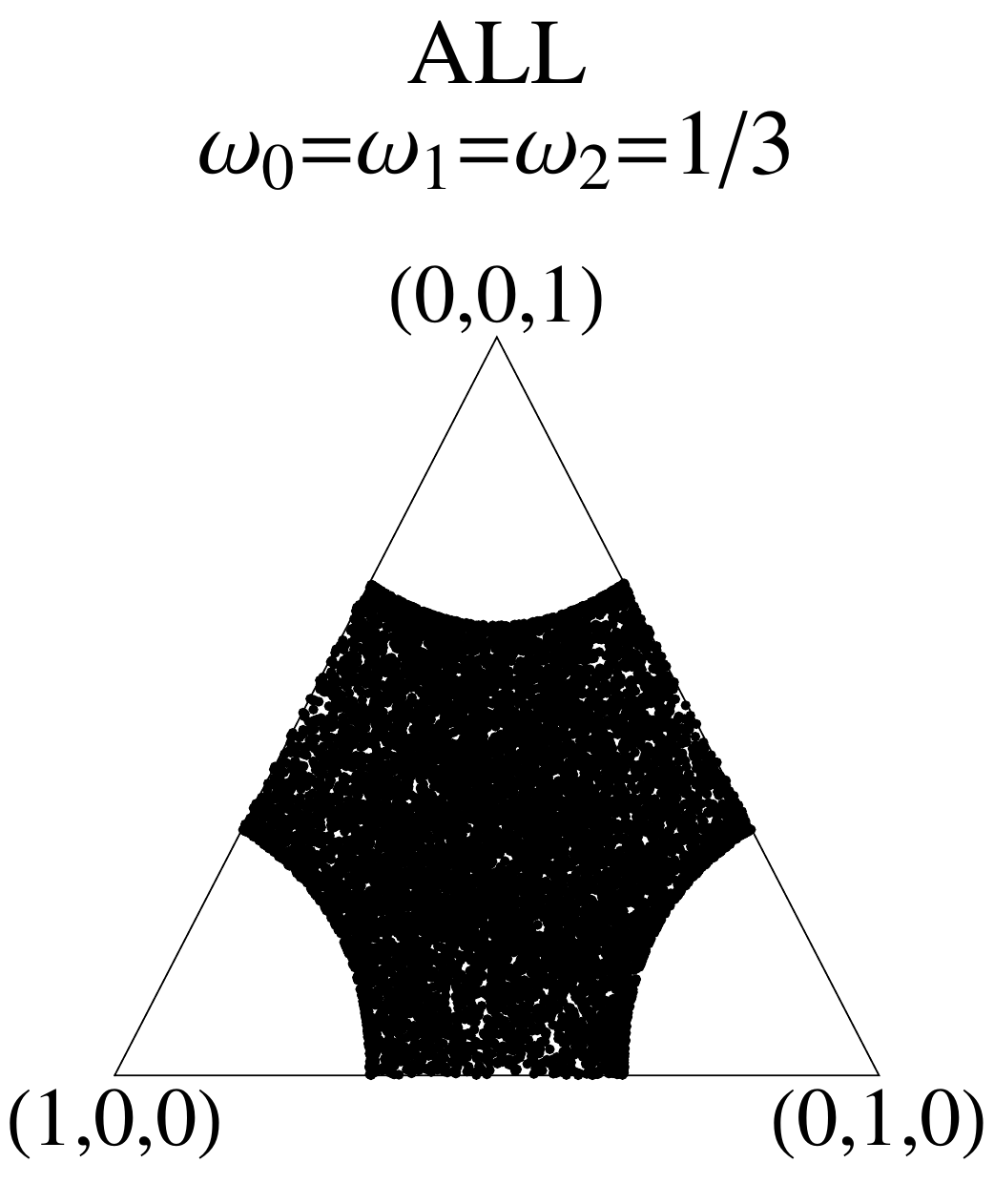}
        \includegraphics[
           height=1.55in,
           width=1.65in]%
          {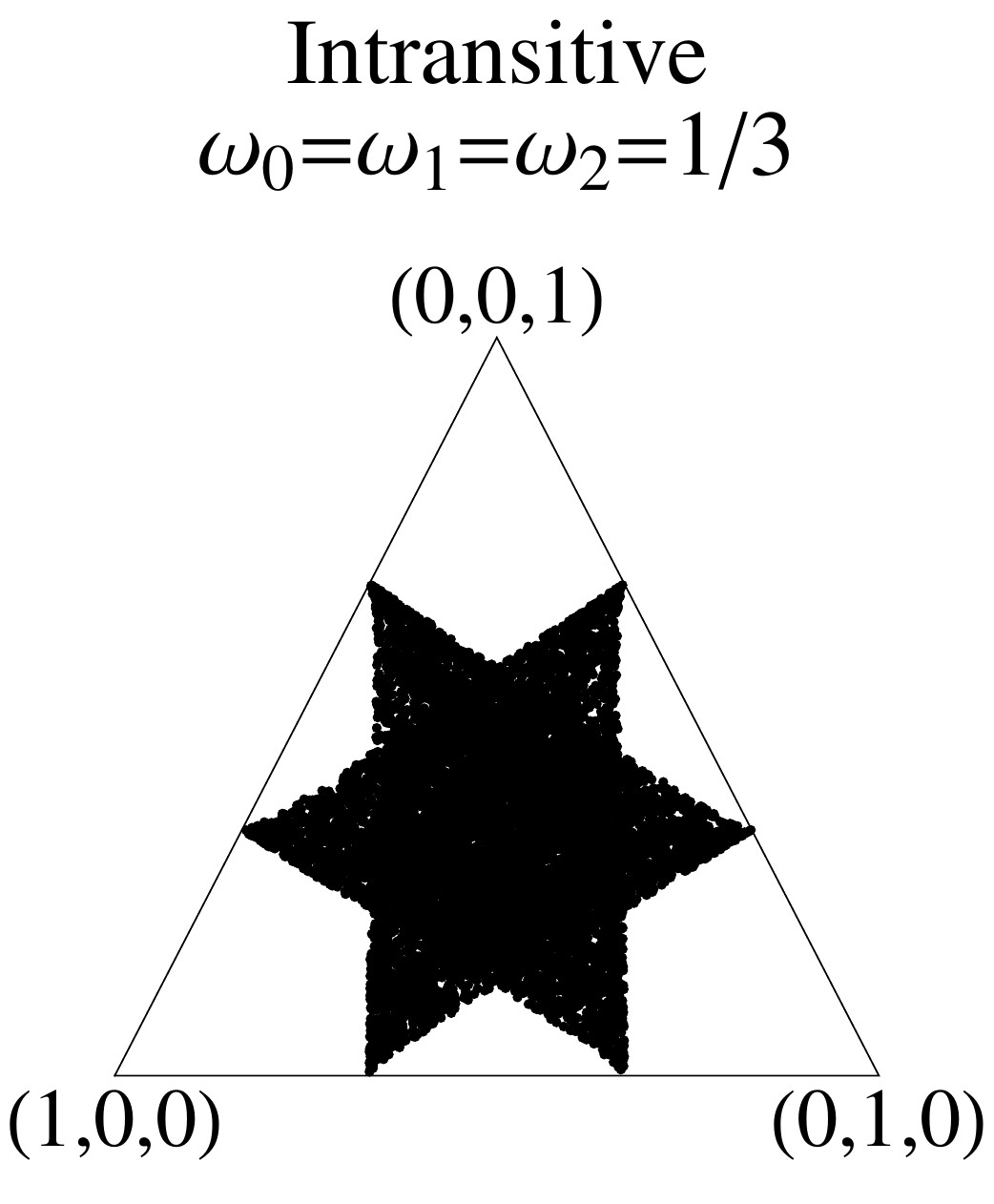}
          \includegraphics[
                       height=1.55in,
                       width=1.65
                       in]%
                      {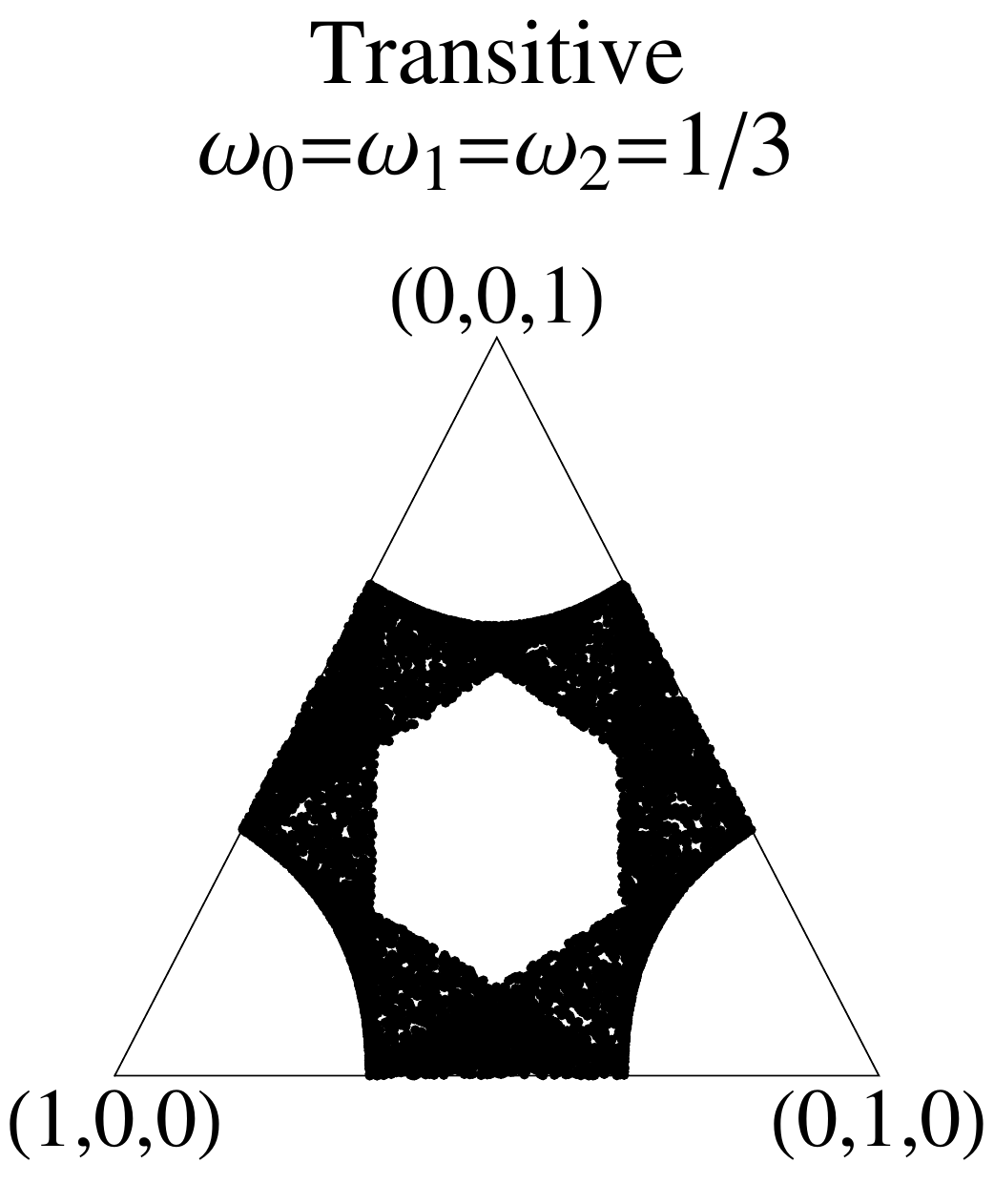}
          }\caption{\textsl{A simplex area for which there exist optimal strategies in quantum model (all, intransitive, transitive) in case $\omega_0=\omega_1=\omega_2=1/3$.}}
           \label{rowne}
\end{figure}

\begin{figure}[htbp]
          \centering{
        \includegraphics[
           height=1.55in,
          width=1.60in]%
          {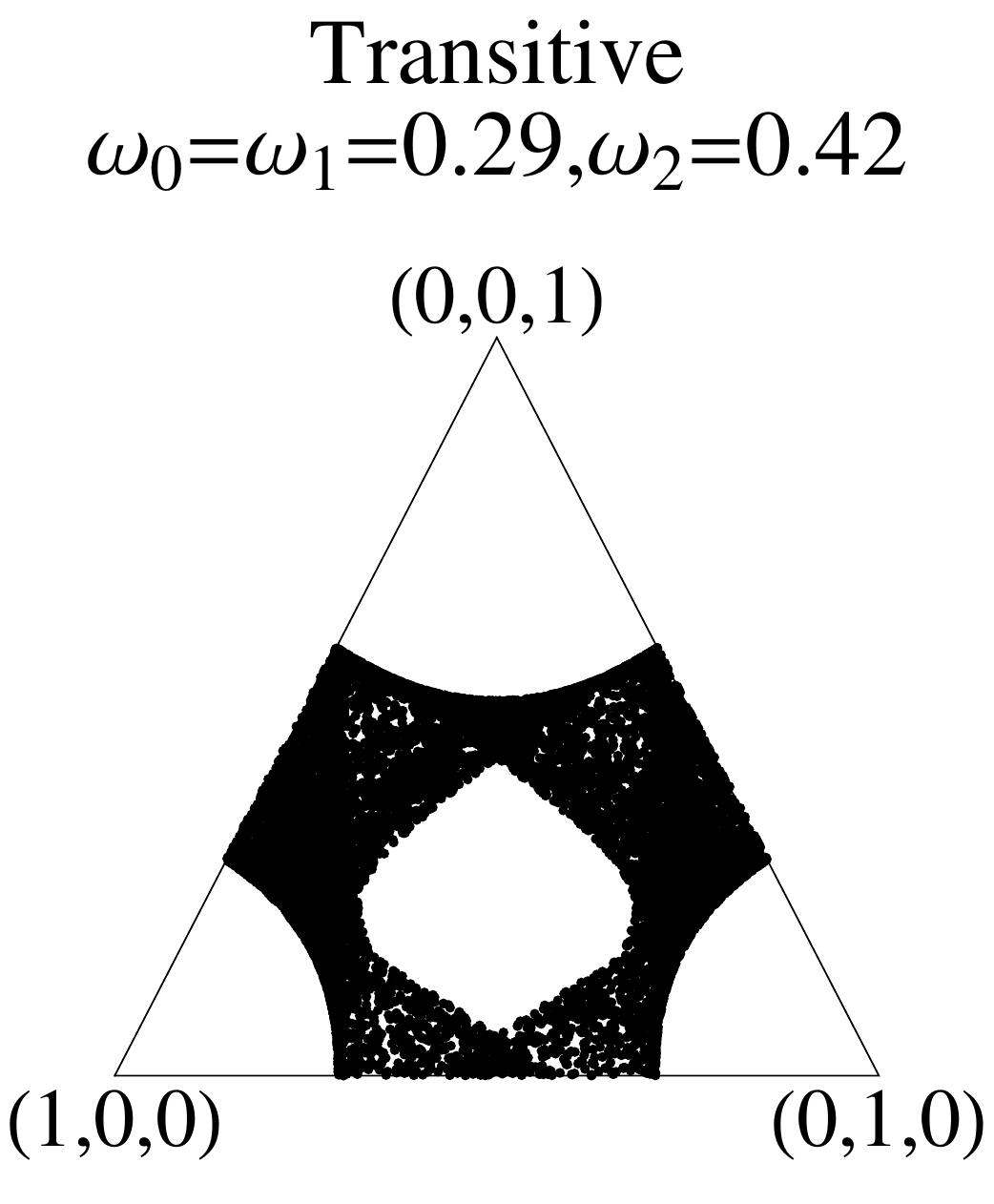}
        \includegraphics[
           height=1.55in,
           width=1.60in]%
          {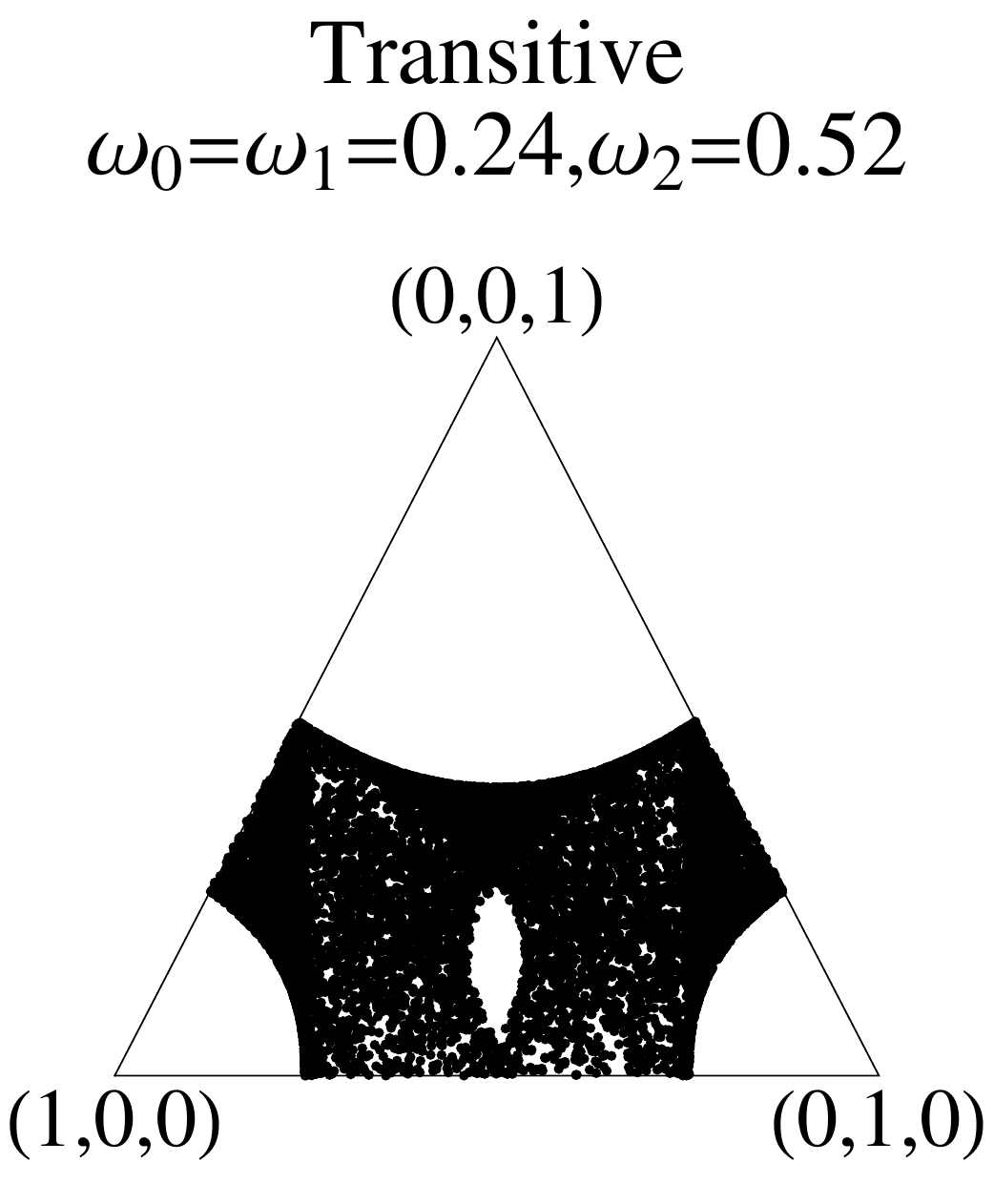}
          \includegraphics[
                     height=1.55in,
                       width=1.60in]%
                       {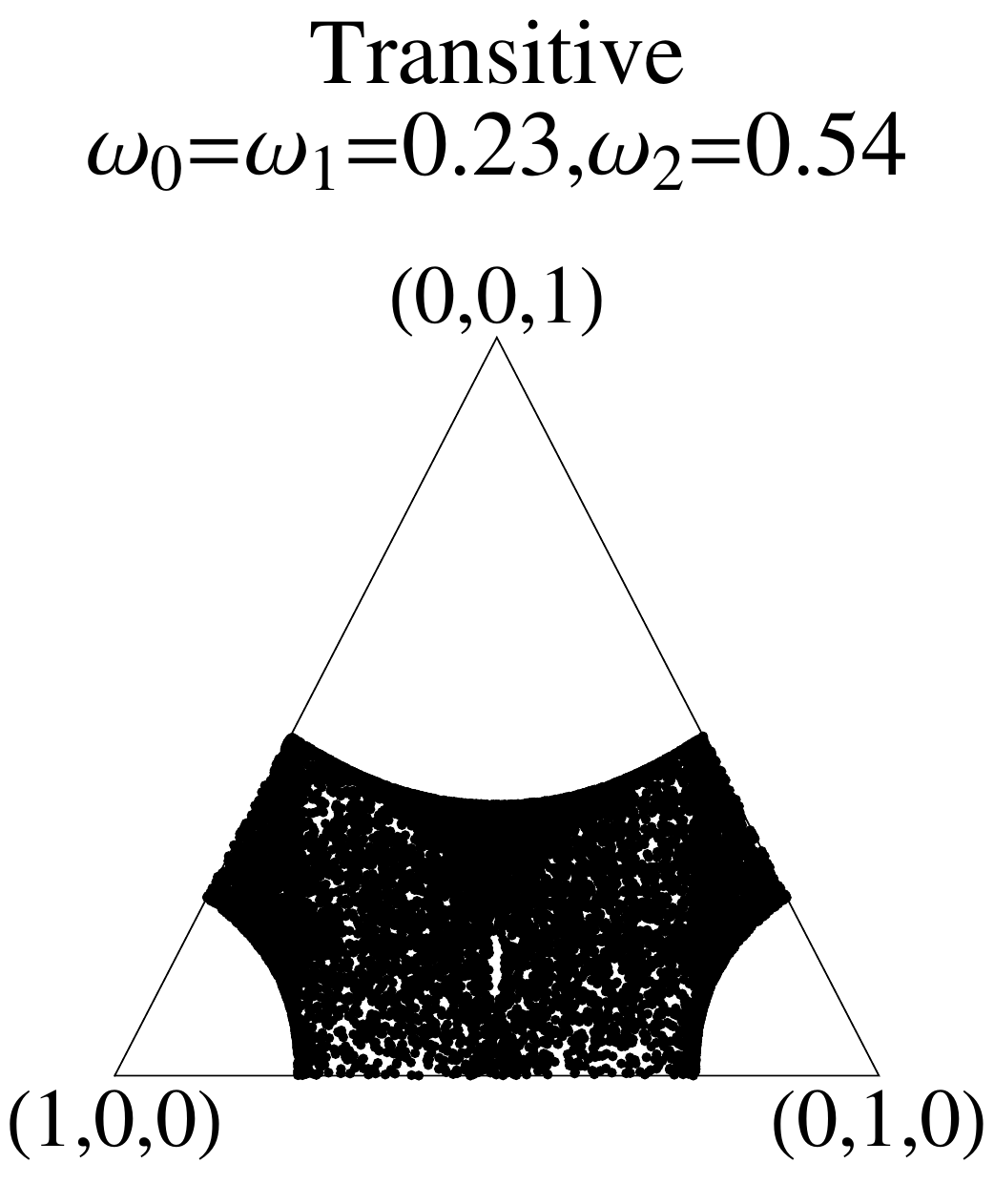}}\caption{\textsl{Decrease of importance of intransitives strategies in quantum model. The dotted area represents the frequencies of $q_m$ for which there is an optimal transitive strategy. A white fusiform slit located in its inside corresponds to the relevant intransitive strategies.}}\label{sp}
\end{figure}

\begin{figure}[htbp]
          \centering{\includegraphics[
           height=1.55in,
           width=1.60in]%
          {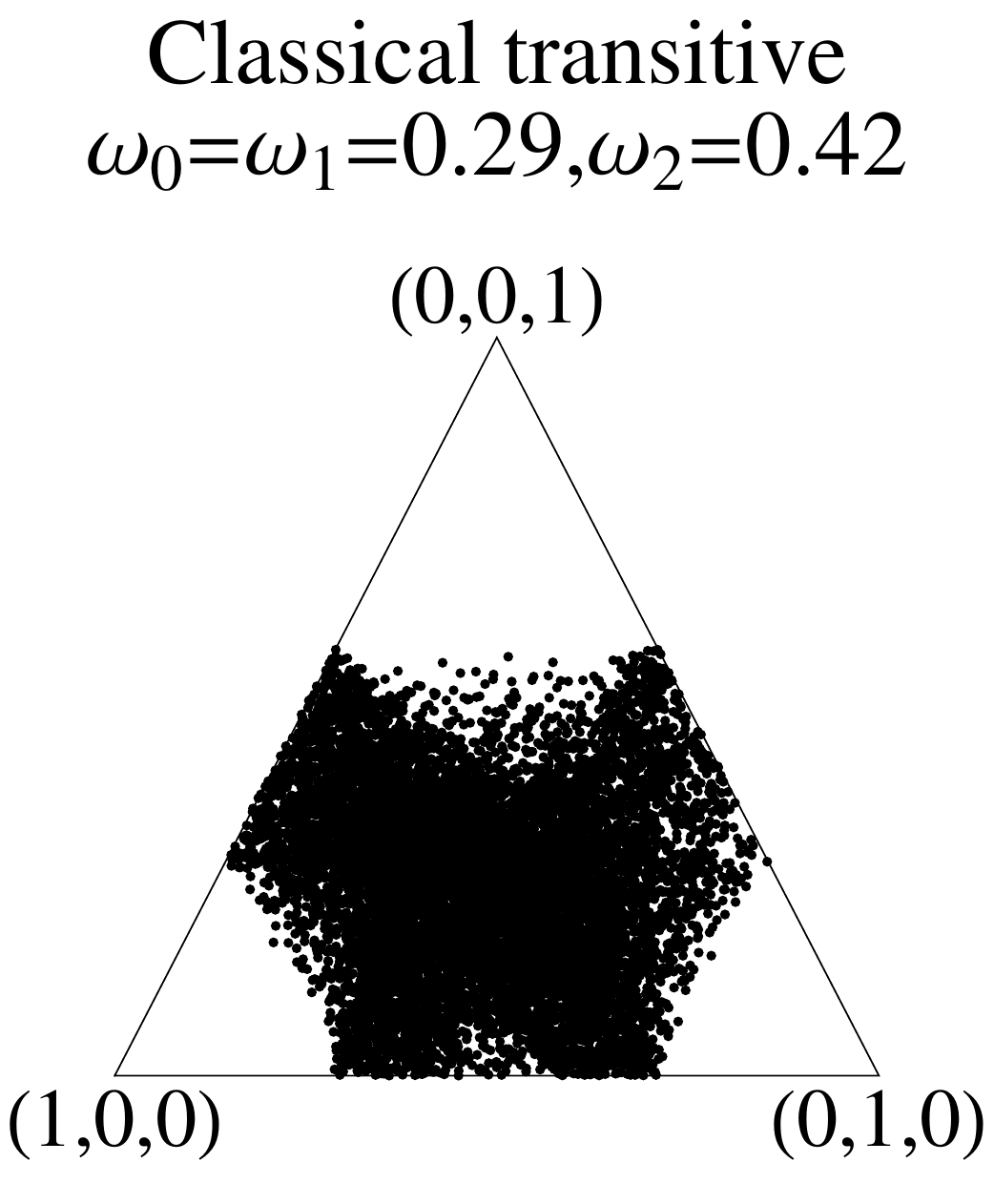}
        \includegraphics[
           height=1.55in,
           width=1.6in]%
          {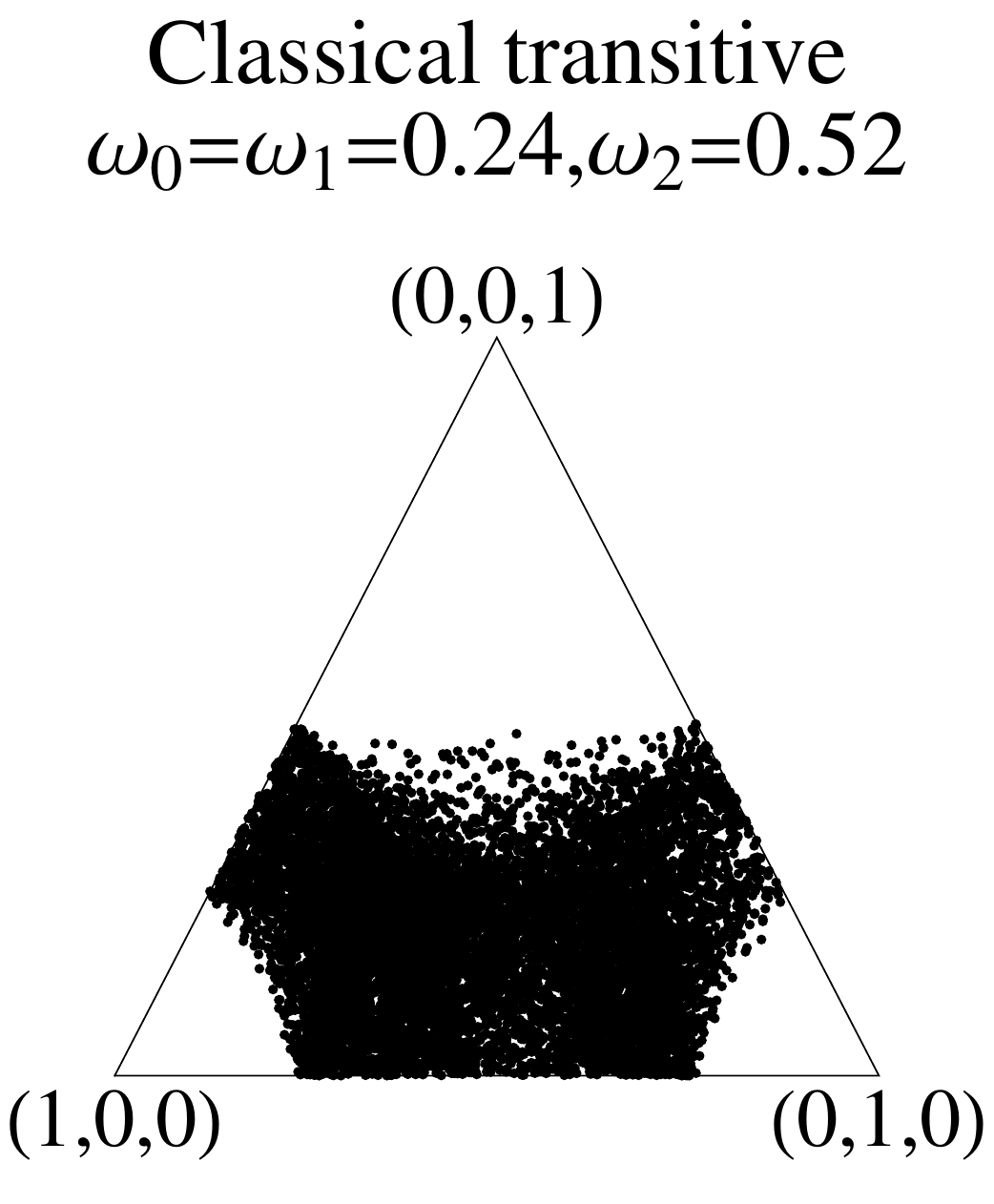}
          \includegraphics[
                       height=1.55in,
                       width=1.6in]%
                       {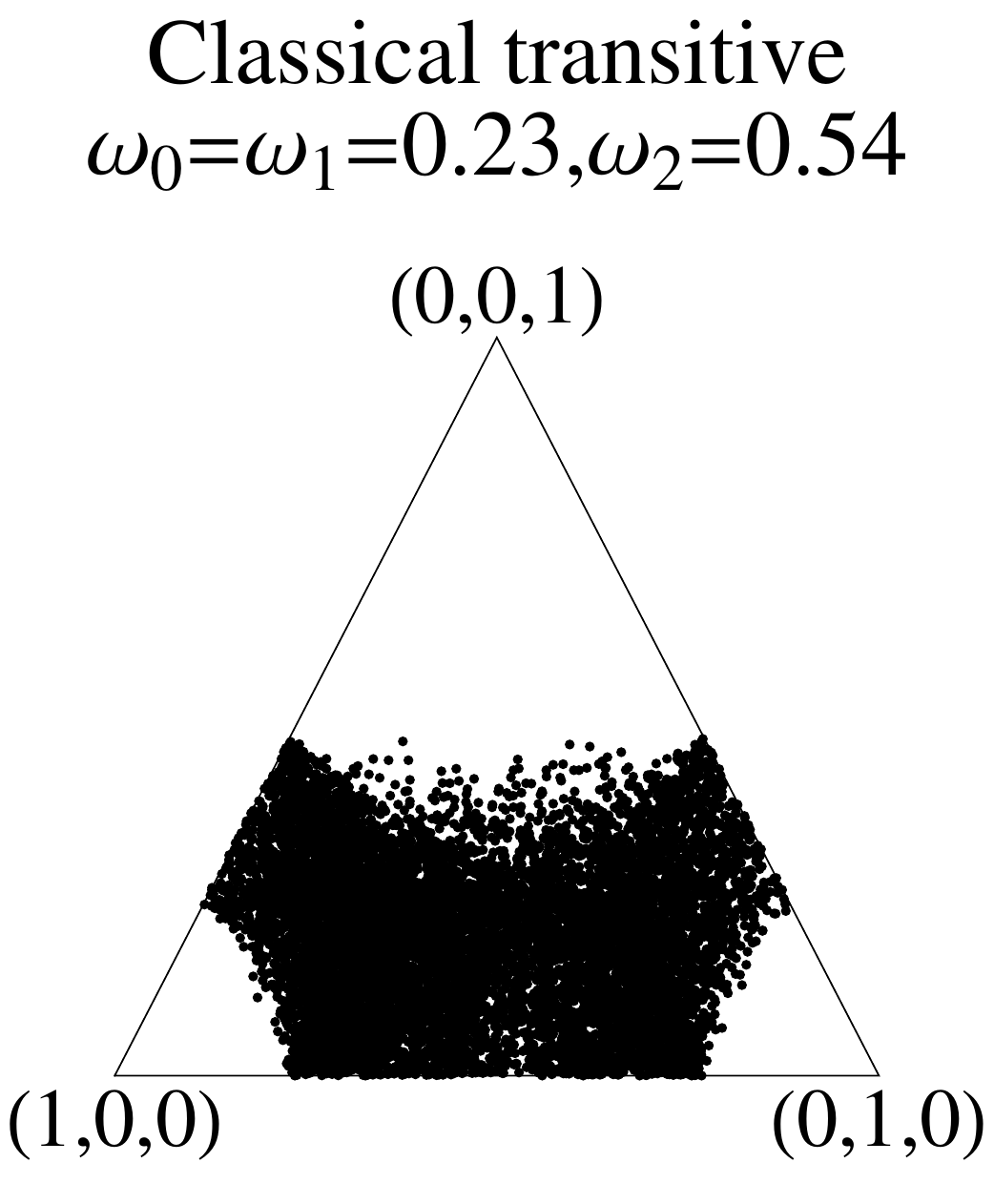}}\caption{\textsl{The dotted area represents the frequencies of $q_m$ for which there is an optimal transitive strategy in classical model.}}
          \label{spC}
\end{figure}

 In this paragraph we will look at the geometrical interpretation of the discussed problem which will allow us to track the changes of the significance of intransitive strategies, if we increase the chance of winning of one of the candidates (in relation to the remaining two). It corresponds to the increase of decisiveness of the electorate --- the collective voter --- that is to some extent the decrease of uncertainty in decision-making.   We present a range of representation $\mathcal{A}_q$ for 10 000 randomly chosen points with the Fubini--Study measure on a sphere $S_2$.\newline
  Fig. \ref{rowne} presents the areas of probability $q_m$ for which  optimal strategies of different types exist (with the assumption that $\omega_0=\omega_1=\omega_2=1/3$ \cite{r2op}).
Candidates have equal chances for winning (none of them has the advantage over the others) which means that the electorate cannot make a decisive choice (is not in favour of a concrete candidate), giving equal support to each of them. It is an exceptional situation which is at the same time extremely interesting from the theoretical point of view, since it illustrates the uncertainty while making decisions when we are not able to conduct the valuation of the assessed goods (in our case the candidates) \cite{r2}. This lack of decisiveness may be a consequence of very similar (practically indiscernible) or incomparable features according to which we make an assessment of the available alternatives (similarly in the case of the abovementioned assessment of fruit tastes which S.M. Ulam tried to conduct). Let us notice that in this case the intransitive strategies (associated exactly with the uncertainty while making decisions) are of great significance. It is clearly observable in the figure presenting transitive strategies. In this case the dotted area ($q_m$ probabilities for which there is an optimal transitive strategy) does not cover the whole area corresponding to optimal strategies of a random type. The non-dotted area in the central part of the figure corresponds to the probabilities $q_m$ for which there is only an optimal intransitive strategy (the relevant intransitive strategy). It is an area of overlapping of two intransitive orders. Let us see how the size of this area is influenced by an increase of support for one of candidates, i.e. the appearance of an election leader.
In this case we will increase the probability $\omega_i$ in relation to the remaining two.    
Fig. \ref{sp} and Fig. \ref{spC} present such a situation (in classical and quantum case) assuming that the probability of winning of candidate  $2$ ($\omega_2$) is increasing and the chances of the other candidates are similar ($\omega_0=\omega_1$). By symmetry, the result is equivalent which ever candidate is chosen to be the leading candidate. We illustrated only the transitive strategies, since in this case the change of importance of intransitive strategies is visible. The figure presents areas which are of interest to us for $\omega_2=0.42$, $\omega_2=0.52$ and $\omega_2=0.54$ respectively. \\
The white fusiform slit (in quantum case Fig. \ref{sp}) in the central part of the triangles corresponds to the area in which two intransitive orders are overlapping (only the relevant intransitive strategy exist, i.e. there is no transitive strategy of identical effect of activity). It means that for a certain distribution of probability $q_m$ of the occurrence of concrete pairs in the second round of elections the voters achieve an established distribution of support ($\omega_0,\omega_1, \omega_2$) only owing to the intransitive strategy.\newline
We can see that along with the increase of the  $\omega_2$ (i.e. the chances for winning of candidate $2$) the white slit in quantum model is becoming smaller  (the importance of intransitive strategies is on the decrease). It will disappear for  $\omega_2\approx0.55$. 
An increase of support for one of the candidates caused the decrease of importance of intransitive strategies (the decrease of the area in which these strategies are relevant). Let us remember that intransitive strategies are associated with an uncertainty and lack of decisiveness. Perhaps by reducing uncertainty we may limit the importance of intransitive strategies.   
\section{Conclusions}
In the discussed model, when the support of a candidate reaches the level of about 55\% (i.e. it exceeds a half which guarantees winning) the intransitive strategies become irrelevant (they still exist but we can always select a transitive strategy of a similar result). It is interesting, since such a level of support means that the voters have made a choice indicating a concrete winner (the candidate who obtained over 50\% of votes). This decisiveness of voters is accompanied by the decrease in the importance of intransitive preferences. Perhaps a certain flaw of the model is the fact that it does not ensue at 51\% already (i.e. the minimum support which is necessary for winning). On the other hand, maybe this minimum advantage does not mean that voters have not decided for a particular candidate. In this case even a minimum decrease of support may change the result of election.\\
The decrease of importance of intransitive orders which accompanies the growth of support for one of the candidates is an interesting property of the quantum game model. This dynamic change cannot be observed in the case of the classical model (see Fig. \ref{spC}) in which for each intransitive strategy we may select a transitive strategy of the same effect. 
In the case of the classical models the relevant strategies (intransitive strategies) cannot occur, since the use of mixed strategies in the description which is necessary in classical models will successfully fill the gaps situated in the surroundings of areas occupied by transitive strategies. Perhaps exactly in the context of the quantum model, the division into transitive and intransitive strategies will allow us to characterise the decisiveness (certainty) or the lack of it (uncertainty) in the process of decision--making. The elementary model presented here constitutes an important example of opportunities provided by non-classical ways of decision--making process description. Extending it with mixed strategies does not remove the effect of occurrence of relevant intransitive preferences \cite{r2op}.
It is also worth emphasising that the classical model (although intuitively clear and intelligible) has however certain flaws which may raise certain reservations. The strategies (conditional probabilities) creating a three--dimensional cube do not have an equal mathematical and information status. In the apex of the dice we have determinist pure strategies, whereas the remaining ones are mixed ones. Strategies may provide different pieces of information (which can be measured with the Boltzmann/Shannon entropy). The quantum model of strategies (pure states) is free from such flaw --- all strategies have an equal informative values (the zero entropy), hence treating them in an equal way (i.e. measuring by means of the Fubini-Study measure) is natural and does not raise any controversy similar of those of the constant measure (the Laplace's principle of insufficient reason \cite{r31}) in the classical model.



\begin{thebibliography}{00}

 \bibitem{r2}  Makowski M 2009 {\it Phys.Lett.A} {\bf 373} 2125
 
 \bibitem{r15}   Boddy L 2000 {\it FEMS Microbiology Ecology} {\bf 31} 185
 
\bibitem{r16}  Shafir S 1994 {\it Animal Behaviour} {\bf 48} 55 

\bibitem{r24}  Arrow K J 1951  {\it Social Choice and Individual Values} (New York: Yale University Press)

\bibitem{r34}   Gardner M  1970  {\it Sci. Amer.} 223 Dec.   

\bibitem{r35} Penney W  1969  {\it Journal of Recreational Mathematics} October

\bibitem{ru} Ulam S M 1976 {\it Adventures of a Mathematician} (New York Scribner) 

\bibitem{r4}  Eisert J,  Wilkens M and  Lewenstein M 1999 {\it Phys. Rev. Lett.}  {\bf 83}  3077

\bibitem{r5} Meyer D 1999 {\it Phys. Rev. Lett.} {\bf 8}  1052

\bibitem{r6} Flitney A P,  Abbott D  2002 {\it Fluct. Noise Lett.}  {\bf 2}  R175

\bibitem{r7}  Piotrowski E W and  S\l adkowski J 2003 {\it Int. J. Theor. Phys.}  {\bf 42} 1089

\bibitem{r26} Miakisz K, Piotrowski E W and  S\l adkowski J 2006 {\it Theoret. Comput. Sci.} {\bf 358}  15 

\bibitem{r8}   Piotrowski E W,  S\l adkowski J 2004 in: C. V. Benton (Ed.), {\it Mathematical Physics Research at the Cutting Edge}  (New York: Nova Science)   pp.247-268.

\bibitem{r9}  Vaidman L 1999 {\it Found. Phys.}  {\bf 29} 615.

\bibitem{r10} Guo H, Zhang J and Koehler G J 2008 {\it Decis. Support Syst.} {\bf 46} 318.

\bibitem{r1} Piotrowski E W and Makowski M 2005 {\it Fluctuat. Noise Lett.} {\bf 5} L85

 \bibitem{r2op}  Makowski M and Piotrowski E W 2006  {\it Phys.Lett.A} {\bf 355} 250
 
 \bibitem{r27}  Wiesner S 1983 {\it Sigact News}  {\bf 15} 78
 
\bibitem{r28}  Paku\l a I,  Piotrowski E W and  S\l adkowski J 2007 {\it Physica A}  {\bf 385} 397

\bibitem{r31}  Dupont P 1977/78 {\it Rend. Sem. Mat. Univ. Politec. Torino} {\bf 36} 125

 \end{thebibliography}
\end{document}